# Economic Optimal Operation of Community Energy Storage Systems in Competitive Energy Markets

Reza Arghandeh[a*], Jeremy Woyak[c], Ahmet Onen[b], Jaesung Jung[b], Robert P. Broadwater[b]

[a] California Institute for Energy and Environment, University of California-Berkeley, Berkeley, CA, USA.
[b] Department of Electrical and Computer Engineering, Virginia Polytechnic Institute and State University, Blacksburg, VA, USA.
[c] Electrical Distribution Design, Inc., Blacksburg, VA, USA.
[*] Corresponding author. Tel.: +1-949-943-5600; e-mail: arghandeh@berkeley.edu
Address: 2089 Addison St, 2nd Floor, Berkeley, CA 94704, USA.

## Abstract

Distributed, controllable energy storage devices offer several benefits to electric power system operation. Three such benefits include reducing peak load, providing standby power, and enhancing power quality. These benefits, however, are only realized during peak load or during an outage, events that are infrequent. This paper presents a means of realizing additional benefits by taking advantage of the fluctuating costs of energy in competitive energy markets. An algorithm for optimal charge/discharge scheduling of community energy storage (CES) devices as well as an analysis of several of the key drivers of the optimization are discussed.

## Keywords

## Nomenclatures

### Symbols

| | |
|---|---|
| $a_t, b_t, c_t$ | : Interpolation coefficients |
| $Ch/Dch$ $Pairprofit_t$ | : CES Charging and Discharging pairs at time t |
| $C_{max}$ | : Maximum CES capacity (kWh) |
| $C_{min}$ | : Minimum CES capacity (kWh) |
| $C_{Rsv\ t}$ | : CES reserve capacity at hour $t$ (kWh) |
| $C_t$ | : CES capacity in time t (kWh) |
| $H_R$ | : Outage support duration (Hours) |
| $K_{config}$ | : Battery cell configuration coefficient |
| $L$ | : Transformer loading |
| $LMP_t$ | : Locational marginal Price in hour t |
| $NT$ | : Number of time points |
| $P_{max}^{Ch}$ | : Maximum charge rate (kW) |
| $P_{max}^{DCh}$ | : Maximum discharge rate (kW) |
| $P_{MaxPri}$ | : Maximum CES power for primary issues (kW) |
| $P_{max}^{Trans\_j}$ | : kVA rating of the transformer j |
| $P_{MinPri}$ | : Minimum CES power for primary issues (kW) |
| $P_t^{CESLoss}$ | : CES loss function (kW) |
| $P_t^{CESout}$ | : Output power of the CES in hour $t$ (kW) |
| $P_t^{FeedLossRed}$ | : Reduction in feeder losses in hour $t$ (kW) |
| $P_t^{Load}$ | : Load in hour $t$ (kW) |
| $R_{cell}$ | : Battery cell internal resistance |
| $R_t^{Ch}$ | : Charging Revenue (Cost) in hour $t$ |
| $R_t^{Dch}$ | : Discharging Revenue (Cost) in hour $t$ |
| $Sch_{profit}$ | : CES Scheduling profit ($) |
| $Sch_{profit}^{opt}$ | : CES Optimal Scheduling profit ($) |
| $SS_{Max}$ | : Maximum iteration step size (kW) |
| $\Delta C_t$ | : Change in stored energy in hour $t$ (kWh) |
| $\Delta C_t^{(i)}$ | : Change in $\Delta C_t$ decided upon in iteration $i$ |

**Acronyms**

| | |
|---|---|
| *AMI* | : Advanced Measurement Infrastructure |
| *CCU* | : CES Control Unit |
| *CES* | : Community Energy Storage System |
| *DCC* | : Distribution Network Control Center |
| *DER* | : Distributed Energy Resources |
| *DESS* | : Distributed Energy Storage Systems |
| *DEW* | : Distributed Engineering Workstation |
| *DMS* | : Distribution Management System |
| *DR* | : Demand Response |
| *ESS* | : Energy Storage Systems |
| *GCU* | : Group CES Control Unit |
| *GHO* | : Gradient-based Heuristic Optimization method |
| *ISM* | : Integrated System Model |
| *LMP* | : Locational marginal Price |
| *PEV* | : Plug-in Electric Vehicle |
| *PBR* | : Performance Based Rates |
| *TOU* | : Time of Use |

# 1. Introduction

Distributed energy storage devices may improve reliability by providing standby power when equipment outages would otherwise force customer interruptions. Additionally, energy storage devices can reduce equipment loading during peak hours, thereby decreasing pre-mature aging in network components [1]. They can also help with renewable energy resource integration into distribution networks. Volt-Var optimization, power quality, frequency regulation, reliability, efficiency, and demand response can all benefit from distributed energy systems [2-5]. These benefits are so great that they sometimes outweigh the high cost of installing the energy storage devices and the communication infrastructure to support them [6].

This paper presents a means of realizing additional benefits from energy storage devices by taking advantage of the fluctuating costs of electricity in competitive energy markets. By combining electricity market information with real-time control of energy storage devices, utilities may enjoy year-round economic benefits from the storage devices, in addition to the occasional benefits mentioned above.

The increasing adoption of intermittent Distributed Energy Resources (DER) into the power grid and technological merit for batteries in recent years brings more attention to energy storage systems (ESS) as viable solutions. Energy storage system integration with renewable sources are discussed in many publications [7-9]. In [7], the authors used a clustering optimization approach to maximize the renewable energy utilization integrated with a pumped storage unit. Authors in [8] explored a large scale battery application for ancillary services in an electricity market. Reference [10] provides a load leveling algorithm with solar power generation and energy storage under a Time of Use (TOU) price scheme. However, the real-time electricity market and the effect of time varying loads were not considered in the demand control algorithm.

Much literature has focused on utility scale energy storage applications (battery capacities more than 1 MW) [11], but few have attempted to realize system wide operational benefits of distributed energy storage systems with battery units with 50 KW and or less capacity. Distributed Energy Storage Systems (DESS) can provide different services for distribution network operators ranging from demand response to power quality issues to peak shaving and renewable resource firming. Moreover, the emergence of microgrids as a special case of network architecture increases the need for DESS [12]. The authors of [13-15] looked at the DESS from the perspective of controlling customer-owned storage devices that integrate with other generation sources. Authors in [16] focused on the DESS application for voltage regulation in the presence of high penetration photovoltaic panels. The customer side of DESS provided voltage regulation in exchange for subsidies from utilities to cover battery costs.

Reference [17] presented a load management approach with substation level energy storage systems for a large load aggregator to determine the electricity price for participation in the day ahead market. A lumped load was considered while distribution grid topology and operational constraints were not considered. In [18] a DESS is used to minimize the forecasting errors associated with DER generation. In [19], the authors integrated DESS into the distribution management system (DMS) controller. However, the DESS is a centralized battery unit to serve the whole substation territory. References [18, 19] and most of the literature related to DMS and distribution network control have proposed a top-down strategy for feeder control starting from the substation. These centralized control approaches need accurate network models and detailed operational constraints for network components to achieve optimal control functionality which is a difficult task [12]. Moreover, energy storage units in those studies are mostly located at the substation.

In distribution networks with DER and DESS sources, the boundaries and operational conditions for each distributed source and the network constraints related to each source need to be included in the control framework. This leads to a distributed control strategy starting from DER and DESS up to the substation. In recent literature, the distributed control approach for DERs is addressed. References [20, 21] present a distributed control system for DESS in distribution networks. However, the DESS control objective is only the feeder loss reduction. The authors in [22] proposed a load management system for residential customers with combined DER and DESS. However, the proposed approach is a single objective optimization to minimize the electricity cost without considering the system's day ahead behavior.

The other school of thought in distributed control strategies for distribution networks is based on Demand Response (DR) programs [23, 24]. DR can play a crucial role in peak shedding and reliability, but there are embedded uncertainties due to DR dependency on customer participation, customer life style, and implementation of Advanced Measurement Infrastructure (AMI) [25].

This paper focuses on the utility owned DESS units installed on residential distribution networks and referred to as a Community Energy Storage (CES) system [26]. The CES term is also addressed in the Department of Energy Smart Grid Recovery Act [27]. The authors of this paper were involved in the CES demonstration project for the State of Michigan, funded by the U.S Department of Energy [28]. The study presented here is based on the actual CES control system design and implementation. The CES unit in this paper is a 25kW Lithium-Ion battery. This paper is not focused on the detailed model of the chemical reactions inside the battery. However, the operational limitations of each CES unit are considered.

From the mathematical point of view, the distributed control approaches have some difficulties with system wide optimal DER operation [29]. This paper proposes a hierarchical control approach for CES scheduling which is a combination of centralized and distributed control approaches.

The other novelty of this paper is in investigating the operation of a utility owned CES fleet in the competitive electricity market, in conjunction with the locational marginal price (LMP). Moreover, the community level energy storage systems placed on the secondary of distribution transformers have not been thoroughly investigated.

Another novelty of this paper is presenting a real-time control scheme that maximizes the revenue attainable with energy storage systems without sacrificing the occasional benefits related to improvements in reliability, efficiency and reduction in peak feeder loading. The proposed multi-objective optimization framework for distributed energy storage is not addressed in available literature. The problem is solved by the Gradient-based Heuristic Optimization method (GHO). The GHO solution uses a combination of trade-offs related to transformer loading, feeder loss, and Locational Marginal Price (LMP) price prediction. The optimal CES schedule in this paper also has real-time and day ahead dimensions, considering the day ahead load forecast and day ahead market price in addition to the real-time circuit measurements and real-time market price.

As previously mentioned, this paper proposes a hierarchical control architecture to carry out the CES optimal control at two levels: the substation level (group CES controller) and the CES unit level. At the substation level, the group CES controller makes optimal decisions and gives optimal commands to the CES units in the distribution network. At the CES level, each CES controller schedules its battery within its local scope and reports its operating conditions and capability to the group CES controller at the substation. The approach for CES scheduling is modular and can be extended to any number of CES units under a substation.

Finally, there has been a lack of physical-based, detailed models in distributed energy resource control and optimization in the literature. The optimization presented in this paper benefits from a detailed distribution network model. The model employed has large numbers of single phase, multi-phase, and unbalanced loads.

The paper is organized as follows: section 2 is a discussion concerning CES infrastructure. Section 3 describes the CES reserve capacity requirements. Section 4 presents the control and optimization problem description. Section 5 addresses assumptions used in the optimization. In sections 6 and 7, case studies and simulation results are discussed.

## 2. Overview of the Community Energy Storage (CES) System

CES units are more flexible than large substation batteries [30]. While the smaller size that is appropriate for secondary distribution does not take advantage of economies of scale, researchers have proposed means of reducing production costs [31-33]. The CES units are connected to the secondary side of distribution transformers to support 120/240 volt circuits. Each transformer serves a small group of houses. References [27, 32] presented guidelines to design community energy storage systems. More important than cost, such placement provides more reliable and secure service to customers. The primary system may be compromised, even to the point of distribution transformer failure. However, the energy storage system may still serve the customer.

Examples of community energy storage system applications for electric vehicles adoption and rooftop solar energy resource penetration are presented in [34, 35] . This paper supposes that the CES units are owned and operated by an electric utility company and thus may be aggregated for system wide optimal operation.

The CES units operate under a hierarchical control system, as illustrated in Figure. 1. Each individual unit is controlled by a local controller, called the CES Control Unit (CCU), which maintains secondary voltage, serves the loads, and handles local issues at each CES unit. The CCU controls the CES unit to charge or discharge based on commands sent from the Group CES Control Unit (GCU), whose set point is given by a control calculation located in the Distribution Control Center (DCC) [34]. In case of islanding or lack of communication, the CCU will control the CES unit without GCU commands.

The GCU algorithm presented here lives in the DCC, where the Locational Marginal Price (LMP) data, forecasted load, and operational alerts, such as storm predictions, are available. Each CCU sends local information, such as the stored energy and transformer loading, back to the GCU.

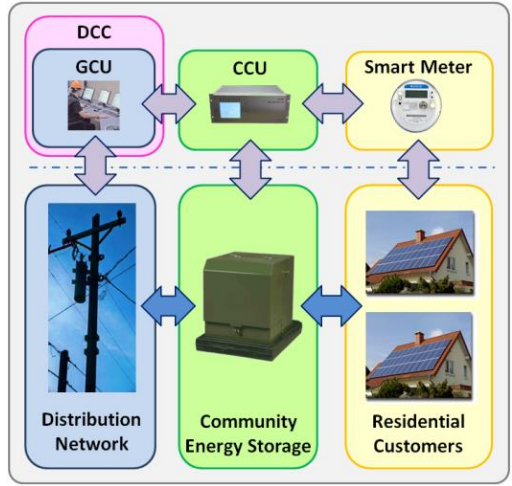

**Figure 1. CES system control layout**

Since the CES units are located on the secondary side of distribution transformers, load management involves, at the secondary level, preventing transformer overloads and low voltages. At the feeder level, it is preventing primary overloads and low voltages. The secondary level constraints (particularly transformer loading) are integrated into the economic scheduling for each CES unit, while the primary level constraints are handled at the feeder level. This involves operating many CES units in a coordinated or aggregated manner. Such constraints are described in more detail in section 4.

The physical-based detailed network model is built in the Distributed Engineering Workstation (DEW) software environment, and is referred to as DEW Integrated System Model (DEW-ISM). Geographical information, component characteristics, load measurements, and supply measurements are included in the model. An ISM offers a graph-based topology iterator framework that facilitates fast computation times for power flow and other calculations on the large scale distribution networks. References [36, 37] provide further explanation about DEW-ISM modeling. In section 6, additional details of the model are presented.

## 3. CES Reserve Capacity Requirements

In the CES control, the reliability benefits take precedence over the energy cost savings. I order to sustain the maximum possible outage durations, the utility can decide to keep the CES units at maximum storage, or above some fixed storage level (*static reserve capacity*). Such lengthy outages may be extremely uncommon, thus resulting in under-utilization of the CES units.

Rather than using a static reserve capacity, the CES units can be managed with a *dynamic reserve capacity*, meaning that the stored energy is kept at a sufficient level to serve an outage for a given duration ($H_R$). For example, suppose that customers on a given feeder may be restored through manual switching efforts which require two hours ($H_R = 2$) to complete. In this case the CES unit only needs to keep enough energy to serve an outage at that transformer for two hours. During peak load conditions the full capacity of the battery may be needed to provide energy for two hours, but during light load conditions the battery may require only a fraction of its maximum capacity to serve the load.

If storm information is available to the GCU, the reserve capacity may be modified in anticipation of longer restoration times. The modeling of the dynamic reserve capacity will be discussed in detail in section 4. The load management constraints and the reserve capacity constraints may be in conflict. Preventing an overload may require violating the dynamic reserve requirement. Here, it is considered better to prevent outages at the present time than to prepare for potential future outages.

# 4. CES Control and Optimization Problem Description

## 4.1. Objective Function for the Optimal CES Scheduling

The objective of the control algorithm is to optimize energy cost savings over time. However, the reliability requirements and component capacities provide constraints for the optimization. When a CES unit charges, the utility incurs the cost of the energy entering into the unit ($R_t^{Ch} < \$0$). When the CES unit discharges, the utility saves the cost of the energy supplied by the unit ($R_t^{Dch} > \$0$). The utility is thus accounting for the difference in energy costs in its own balance sheet as a savings.

In addition to the energy going into or out of the CES units, the utility may also count the change in feeder losses in its cost savings calculation. That is, when the CES unit is charging, the load has increased, and there will be greater losses in the feeder due to the increased current. Similarly, when the CES unit discharges, the losses on the feeder decrease and the CES control may take "credit" for such cost savings. Combined with the cost savings due directly to the CES output energy, this net cost savings will be termed "revenue" when taken separately ($R_t^t$, $R_t^{Dch}$), and "operating profit" or "profit" when combined (*Ch/Dch Pairprofit, Sch$_{profit}$*).

Since the goal is to maximize profit by taking advantage of CES stored energy, the two primary drivers of the optimization algorithm are the LMP prediction (cost of energy) and the load forecast.

In regulated electric energy markets, the Locational Marginal Price (LMP) is computed in real-time based on bids from energy producers, losses and line congestion. These prices are called the real-time LMPs and represent the incremental cost to supply load to a given region at a given time [38]. In addition to the real-time LMP market, there is a day-ahead LMP market, wherein energy producers bid their expected costs one day ahead [39]. In order to determine the optimal charging and discharging schedule for the CES units, the price of energy at future hours is needed. Since the real-time LMP price is not known in advance, the utility may use either the day-ahead LMP or, if it has a better prediction of its own costs based on its generators, it may use its own LMP forecast. The simulation reported here uses the day-ahead LMP price and compares the profit with that attainable from an ideal prediction (ie, a prediction that exactly matches the LMP prices that occur over the next 24 hours).

The distribution network load forecast is the other primary driver of the economic optimization. The load forecasting provides the time-varying load estimate for each distribution transformer for the next 24 hours. The distribution transformer loading is necessary for each battery's day ahead schedule. The 24-hour load forecast for each transformer is based on a "load research statistics" load estimation method that is developed in [40, 41]. The real-time demand at each distribution transformer with a CES unit is metered locally and provided to the GCU for use in the optimization algorithm to refine the load forecast. Weather data may also be used to refine the load forecast [42].

In the real-time application, the optimization algorithm first calculates the optimal charge/discharge schedule for each CES unit for the next 24 hours, and then issues the commands through the GCU for the first hour's optimum operation. Each CES unit is scheduled independently using the same formulation, but the constraints will be set up differently as the corresponding measurements differ. Mathematically, the objective function then has 24 independent variables—the kW output at each hour—and one dependent variable (the total profit).

The objective function to be optimized is given in (1):

$$Sch_{profit} = max \sum_{t=0}^{23} LMP_t * (P_t^{CESout} + P_t^{FeedLossRal}) \qquad (1)$$

, where $P_t^{CESout}$ is the CES power output between the $t^{th}$ and $t^{th-1}$ hours. The sign convention used here associates a positive $P_t^{CESout}$ with discharging and a negative $P_t^{CESout}$ with charging. Similarly, $P_t^{FeedLossRed}$ is negative when the losses increase.

As noted, the total feeder losses do not need to be calculated for each output; rather, only the change in feeder losses needs to be calculated. Since feeder losses are quadratic for radial distribution systems, the change in losses due to any given CES unit's output are computed by interpolating a quadratic polynomial defined by three points relating CES output to feeder losses.

$$P_t^{FeedLossRed} = a_t \times (P_t^{CESout})^2 + b_t \times P_t^{CESout} + c_t \qquad (2)$$

The coefficients $a_t$, $b_t$, and $c_t$ are calculated in advance using power flow calculations on the feeder for three different output levels. Thus, rather than running a power flow computation on the entire feeder at every iteration in the scheduling algorithm, only three power flow runs per battery are required.

## 4.2. Equality and Inequality Constraints for the Optimal CES Scheduling

In the optimization problem system, constraints are related to the CES unit properties and the distribution network operational properties. CES properties include efficiency, capacity, state of charge, charging rate, discharging rate, and maximum capacity. The distribution network operation properties include the voltage level, transformer capacity, and the reliability requirement.

Since the CES outputs are assumed to be constant for the entire hour, the conversion of output power (kW) to stored energy (kWh) involves only the identity function. However, the CES units do not operate at perfect efficiency and the losses incurred during both charging and discharging must be considered. Equation (3) relates the actual CES output power to the change in stored energy $\Delta C_t$, where the internal losses $P_t^{CESLoss}$ are a function of the CES internal resistance which is provided by the battery manufacturer.

$$\left| \Delta C_t \right| = \left| P_t^{CESout} + P_t^{CESLoss} \left( P_t^{CESout} \right) \right| \qquad (3)$$

The model used to compute the CES internal losses consists of a constant voltage source in series with a resistor, as given by

$$P_t^{CESLoss} = \frac{R_{CES}(temp)}{\left( V_{cell} \right)^2} * (P_t^{CESout})^2 \qquad (4)$$

$V_{cell}$ is the battery cell output voltage. $R_{CES}(temp)$ is the resistance of a CES unit, and is a function of the unit temperature, battery cell number, and the battery cell configuration in the CES unit.

As mentioned in section 3, the operation of the CES units includes both secondary (local) and primary (fleet) constraints. Each CES unit has to first satisfy local constraints, such as preventing a distribution transformer overload, while serving with other CES units in the fleet to meet primary constraints, such as preventing an overload at the substation exit cable. Even before the secondary power system constraints are imposed upon the schedule, the physical constraints of the CES units must be met, per the manufacturer's specifications. These constraints include power output (kW) and stored energy (kWh).

The output power constraint in each hour is given in (5), where $P_{max}^{Ch}$ and $P_{max}^{DCh}$ are maximum charging and discharging rates, respectively. The negative sign in front of $P_{max}^{Ch}$ reflects the sign convention used here, where charging is "negative output."

$$-P_{max}^{Ch} \leq P_t^{CESout} \leq P_{max}^{DCh} \qquad \forall 0 \leq t \leq 23 \qquad (5)$$

The CES stored energy constraints are modeled in (6), where $C_{min}$ and $C_{max}$ are the minimum and maximum levels of energy that can be stored by the CES unit, respectively, per the manufacturer's operating recommendations. $C_t$ is the battery capacity at time $t$. The $C_{min}$ limit prevents the battery from being fully discharged.

$$C_{min} \leq \left| C_t \right| \leq C_{max} \qquad \forall 0 \leq t \leq 23 \qquad (6)$$

Once the physical CES constraints have been computed, the power system primary and secondary constraints are computed. The local loading measurements and load forecast are first used to set bounds on the CES unit outputs so as to prevent transformer overloads. These constraints take the form shown in (7).

$$\left| P_t^{Load} - P_t^{CESout} \right| \leq P_{max}^{Trans\_j} \qquad (7)$$

, where $P_t^{Load}$ is load (kW) at time $t$ under the transformer $j$. $P_{max}^{Trans\_j}$ may be set equal to the kVA rating of the transformer. Although the reactive component of the load will add to the total kVA loading on the transformer, the CES unit may supply power to the system through an inverter, which may supply VARs up to a certain limit. Since transformer ratings are not firm ratings (a transformer may be slightly overloaded with minimal impact on reliability and transformer life), this approximation was deemed appropriate.

Together with the availability of stored energy in the CES unit, the transformer loading constraint is used to calculate the CES availability for primary-level control. Power flow analysis is then used to check for primary-level overloads and low voltages. If such primary-level problems exist, the CES fleet will be used to attempt to alleviate those problems. These changes create additional constraints, as described in (8), that are placed upon the economic optimization.

$$P_{MinPri} \leq P_t^{CESout} \leq P_{MaxPri} \ \forall 0 \leq t \leq 23 \qquad (8)$$

,where $P_{MinPri}$ and $P_{MaxPri}$ are minimum and maximum CES power for primary level issues. While the formulation of this algorithm focuses on individual CES units independent of other CES units, additional CES units may greatly affect the participation each unit plays in resolving primary-level constraints. Each CES schedule is different because each transformer has a different loading pattern.

As mentioned in section 3, the reserve capacity (kWhr) may be either static or dynamic. If static, the user specifies $C_{Rsv\,t0}$ directly. If dynamic, then the reserve capacity may be calculated by

$$C_{Rsv_{t_0}} = \int_{t_0}^{t_0 + H_R} Ld(t)dt \qquad (9)$$

,where $H_R$ is the number of outage support hours as explained in section 3. The loading on the transformer $Ld(t)$ is based on typical customer load curves [40] modified based on recent measurements and available load forecast information at that location.

The reserve energy is therefore constrained by

$$-C_{Rsv\,t} \leq \Delta C_t \leq C_{max} \qquad \forall 0 \leq t \leq 23 \qquad (10)$$

## 4.3. Gradient-based Heuristic Optimization Solution

The optimization solution for equation (1) first identifies the schedule with minimum charging and discharging that satisfies the various constraints at each hour. If there are any infeasible constraints the reserve capacity constraint will be violated before the transformer loading constraint is violated. Then, starting from the initial schedule, the algorithm proceeds to add equal amounts (kWh) of charging and discharging at each iteration, moving toward an optimal schedule.

Mathematically the Gradient-based Heuristic Optimization (GHO) algorithm is viewed as a heuristic version of the gradient method for solving optimization problems. The "gradient" is the marginal cost of increasing a unit of energy stored or released during a given hour. This unit of energy $\Delta C_t^{(i)}$ is called the step size. The maximum step size $SS_{Max}$ for iterative optimization solution is chosen by the user, based on the level of accuracy required and the amount of execution time available where

$$\Delta C_t^{(i)} < SS_{Max} \qquad \forall \ 0 < t < 23 \qquad (11)$$

The revenue $R_t$ corresponding to a given unit of energy at time $t$ is expressed in (12). Note that $R_t$ will be negative when charging and positive when discharging.

$$R_t = LMP_t \times (P_t^{CESout} + P_t^{FeedLossRed}) \qquad (12)$$

At each iteration charging is added to only one hour and discharging is added to another hour, based on the charge/discharge pair with the maximum profit as given by

$$Ch/DchPair_{\text{profit}} = \max(R_t^{Ch} + R_t^{DCh})$$
$$\forall 0 \le t \le 23 \tag{13}$$

In (13) charging and discharging do not happen at the same step. The charging or discharging at each iteration maintain a consistent time-integrated amount of energy stored inside the CES unit. Due to internal losses in the battery and inverter, the power output and input seen external to the CES unit will be slightly different.

When a constraint is reached in (10), the actual step size at that time slot $\Delta C_t$ may be smaller than $SS_{Max}$, but that time slot is then no longer considered in future iterations. Because of the "greedy" characteristic of the GHO algorithm, the maximum number of steps required to find the optimal schedule for each CES unit can be calculated by

$$MaxSteps = \frac{max(P_{max}^{Ch}, P_{max}^{DCh})}{(SS_{Max}) * (NT)} \tag{14}$$

,where $NT$ is the number of time points. As long as both loss functions, the CES internal loss and the feeder loss, are convex in all dimensions (as, for example, when the CES losses are modeled by a quadratic function dependent on a modeled internal resistance), the objective function as a whole is convex. Therefore, the algorithm converges to the optimal solution as the step size is reduced, where reasoning similar to the proof in [43] may be used.

$$\lim_{SS_{max} \to 0} (Sch_{profit}^{Opt} - Sch_{profit}) = 0 \tag{15}$$

Finally, the algorithm will only schedule charging and discharging when it can make greater than a certain minimum profit per kWh, specified as *Minprofit*, on the charge/discharge cycle as

$$\frac{Ch/Dch\,Pairprofit_t}{2 * \Delta C_t} > Minprofit \quad \forall\ 0 < t < 23 \tag{16}$$

Since charging and discharging activity reduces the battery's life, the utility will set the minimum profit margin based on the CES unit's cost and life cycle information from the battery manufacturer. Thus, the algorithm will either stop when all time slots have been scheduled or when the maximum charge/discharge pair profit is less than or equal to the minimum profit (based on battery life cycle data from the manufacturer). Figure 2 depicts the GHO algorithm for the CES economic optimal scheduling problem. The equations that are solved in each step are indicated in Figure 2.

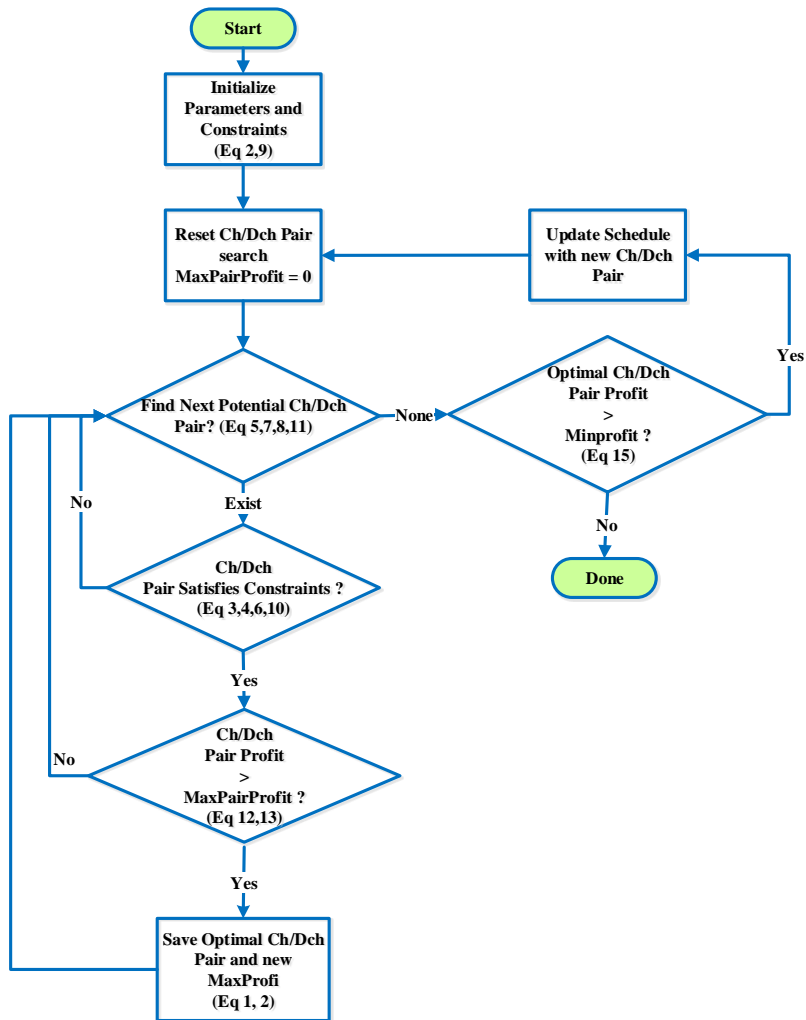

**Figure 2. Gradient-based heuristic optimization algorithm for the CES optimal scheduling**

## 5. Assumptions for the Optimization Solution

The optimization algorithm should be run as frequently as the LMP prices or demand measurements change. Essentially, the 24-hour schedule is only used to decide what to do at the present time. The algorithm may be run again in another five or ten minutes to revise the current operation as the loading and energy costs have changed. It is worthwhile to mention that the algorithm average computational time in the DEW software environment is less than 5 minutes, which is based on numerous simulations. Because of how quickly load and LMP prices change, it is not feasible to operate the CES units based on information that is almost 24 hours old. It is also not feasible to operate based on information that is only 15 minutes old, as the LMP price is subject to occasional spikes. It is a continuous process, and when the "end of the day" arrives, the algorithm is already looking ahead to the end of the next day and adjusting the schedule accordingly.

Not only does the LMP price change rapidly, but also the load forecasting may be inaccurate. Again, by re-running the CES control algorithm every five to ten minutes with an updated load forecast, the CES units will be able to respond more quickly to changes in load.

While the algorithm described in section 4 may be used with many varieties of storage technologies, rated outputs, storage capacities, and losses, the simulation presented here considers CES units with Lithium-Ion cells as recommended in [44]. They are 50kWh energy storage and have the maximum discharge rate of 25kW. Lithium-Ion battery self-discharge is neglected in this paper. Lithium-Ion batteries are a commonly used technology in commercial distributed energy storage solutions because of their high energy and power density, long life cycle and safety issues [45, 46].

The load research data, customer billing information, and primary system measurements are applied to model time varying conditions [47].

The "profit" term in the following sections refers to the CES unit's operational cost savings. The results presented focus on the CES optimal economic operation, where the capital cost and installation cost are not considered in operational profit.

## 6. Simulation Results and Trade-Off Analysis

### 6.1. LMP Prediction Accuracy

Not surprisingly, the accuracy of the price forecast has a significant impact on the cost savings potential. The real-time price will often deviate from the day-ahead bids. In such cases, if the CES unit is charged when the price dips or discharged when the price rises, it will be unable to take advantage of these price fluctuations. Table 1 shows the total profit for one month's operation of one CES unit calculated using the day-ahead prices as well as the total profit that could have been achieved if the real-time prices were known 24-hours in advance (an "ideal prediction"). Thus, there is an economic incentive for utilities to revise the LMP forecast [48] when possible. It is worth mentioning that the day ahead and the real time LMP prices are dictated by Independent System Operators (ISO) to utilities. Utilities do not determine the LMP prices. The difference between day ahead and real time LMP is due to unforeseen changes in generation capacity, transmission constraints, or load. The impact of the load profile change on the LMP price is ignored in the scope of this work because the capacity of the batteries is extremely small when compared with the total load for the locational node in the electricity market.

The energy cost for a representative distribution transformer without a CES unit was calculated to be $284 based on a 0.12 $/kWh electricity tariff in the Detroit, MI area. If a CES unit is added downstream of the representative distribution transformer, the optimal CES operation profit was calculated to be 35% and 65%, respectively, for day-ahead LMP forecast and ideal LMP forecast cases. See Table 1.

**Table 1. LMP Accuracy Analysis for Monthly Profit of Each CES Unit**

| | CES OPERATION PROFIT | CES PROFIT / RETAIL PRICE* |
|---|---|---|
| Day-Ahead LMP | $101 | 0.355 |
| Ideal Prediction | $185 | 0.651 |

\* Electricity retail price for the case study transformer without CES during a month is $284 based on 0.12 $/kWh residential tariff.

### 6.2. Transformer Loading and Reserve Capacity

When the load is higher, more energy must be saved in case of an outage, so there is lower availability for participation in the energy market. It is resulting in smaller profits. On the other hand, when the system load is higher, the energy prices tend to be higher, resulting in higher profits. The simulation presented here uses the month of July, when both the load and the prices tend to be highest.

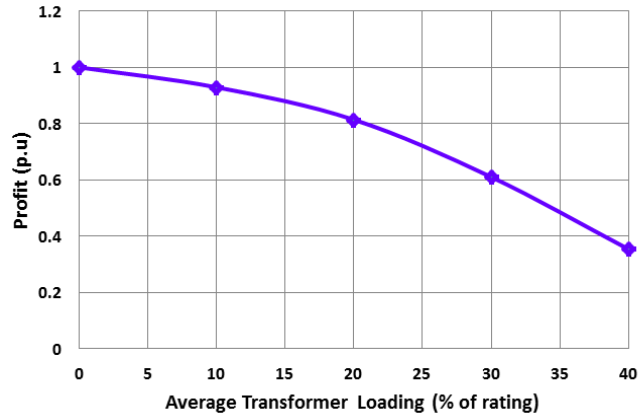

**Figure 3. CES operation profit vs. distribution transformer loading.**

Since a higher load means less energy is available for market participation, a greater profit can be realized by placing the CES units on more lightly loaded transformers. Figure. 3 illustrates this tradeoff, plotting the profit (cost savings) against the transformer loading. However, the CES placement does not only depend on the transformer loading. Sometimes utilities place CES in areas with more outages. With a heavily loaded transformer, the utility is trading off operating profit for improved reliability.

The dynamic reserve capacity is dependent not only on the load but also on the number of hours' worth of load that the CES unit is expected to be able to serve in case of an outage. Figure 4 illustrates the tradeoff between the profit potential of the CES unit and the number of hours of reserve capacity. Utilities must decide how much reliability they are willing to risk to realize increased savings. Utilities with performance based rates (PBR) could use a plot, like the one in Figure 4, to identify where the tradeoff between profit and reliability matches their PBR [49].

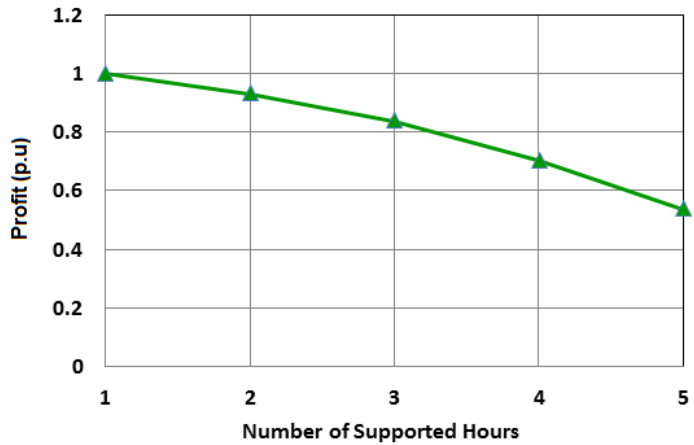

**Figure 4. CES operational profit vs. number of hours for supporting loads following an outage.**

For comparison, a static reserve margin is also assessed. Figure 5 compares the use of a static reserve capacity with a dynamic reserve capacity. The static reserve capacity is fixed at the maximum capacity used by the dynamic reserve capacity.

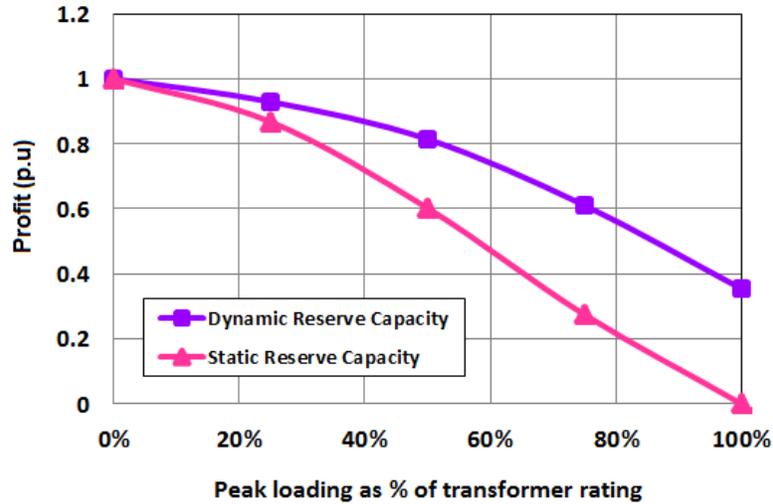

**Figure 5. CES dynamic and static reserve operational benefits**

The static reserve capacity benefit is 39% less than the dynamic reserve capacity method at 80% transformer loading. In this curve a residential load is used. Customers with different load shapes would see different variations between dynamic and static reserve capacity.

Figure 6 shows the tradeoff between profit and fixed (static) reserve capacity. While a dynamic reserve margin offers a greater potential cost savings, computing the dynamic reserve margin depends heavily on accurate load forecasting. A utility with a poor load forecasting capability may be limited to using a static reserve capacity. Similarly, a customer with a constant load profile has a static reserve capacity.

Alternatively, Figure 6 can be understood as depicting the tradeoff between energy storage size and cost savings, as opposed to reliability and cost savings.

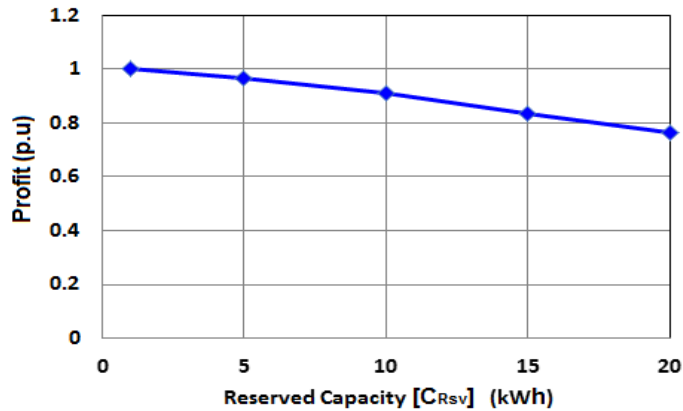

**Figure 6. CES Operational profit vs. different fixed reserved capacity of a CES**

The cost savings potential is not only limited by the reserve capacity (stored energy), but also by the output power required to prevent overloads on a transformer.

Figure 7 depicts the effects of the transformer loading constraint on the CES charging/discharging schedule. In Figure 7 adoption of Plug-in Electric Vehicle (PEV) loads at 30% of the average transformer load creates an overload between 5 PM and 9 PM on the distribution transformer. The cost savings potential is not only limited by the reserve capacity (stored energy), but also by the output power required to prevent overloads on a transformer. The dotted line in Figure 7 is the distribution transformer nominal capacity (50 kW), the dashed line is load with 30% PEV adoption, and the solid line is the load with 30% PEV adoption and CES.

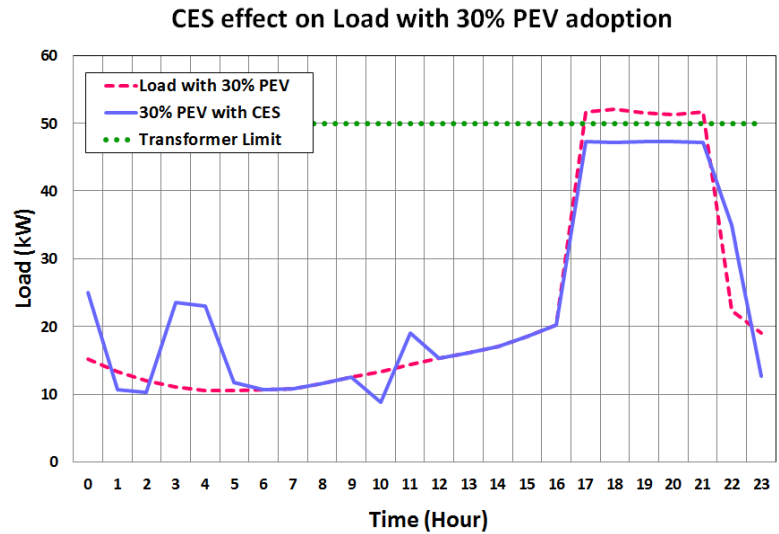

**Figure 7. Load profiles considering 30% PEV adoption with and without CES**

The CES constraints ensure that the battery retains sufficient charge to prevent a transformer overload during the entire four-hour period in which the PEVs are being charged [34].

### 6.3. Impact of the Feeder Losses

The significance of feeder losses is presented in Table 2. The profit is calculated for a single month for a single CES unit that is far from the substation on a moderately loaded feeder. Two approaches are used to schedule the CES charging/discharging. The first approach includes feeder loss reduction as a part of the objective function (see equation (1)), as well as the internal battery losses. The second approach ignores the feeder losses in optimization, but still incorporates internal battery losses. Note that an ideal LMP "prediction" was used in calculating the results shown in Table 2.

**Table 2. Monthly Profit for a CES Unit**

| Objective Function | Profit ($) | Computational Time (Sec) |
|---|---|---|
| Lmp Price + Loss Function | 185 | 337 |
| Lmp Price | 182 | 90 |

Table 2 shows that the feeder losses have a very small impact on the total cost savings attainable by the optimization. The computational time to calculate feeder losses is almost three times the total computational time for maximizing the objective function. The quality of the price forecast and the internal battery losses both outweigh the feeder losses in determining the optimal schedule.

### 6.4. Impact of the Optimization Step Size

Table 3 shows the effects of changing the optimization step size on the profit achieved by the optimization algorithm in section 4 (see equation 11), where the optimization is performed for an entire month for the LMP price case (second case in Table 2). Since most of the cost savings comes from hours with unusually high or low LMP prices, changing the step size has relatively little impact on the cost savings realized by the algorithm. But, it has a significant impact on the computational time. Note that the actual step size used in each iteration depends also on the constraints – even if the step size is 5 kWh. If the battery only has 2 kWh of remaining stored energy in a given hour, the step size effectively becomes 2 kWh for that iteration.

**Table 3. CES Optimization Step Size**

| Step Size (% Of Max Battery Capacity) | Computational Time For One Month (Sec) | Profit For One Month ($) |
|---|---|---|
| 2 | 90 | 101 |
| 5 | 65 | 101 |
| 10 | 57 | 101 |
| 20 | 55 | 100 |

### 6.5. Aggregated Response of the CES Fleet

The DEW-ISM model used for the case study represents an actual 13.2kV residential feeder in the Michigan with a 1769 kVA annual peak load. It has single phase and multiphase unbalanced loads. 20 CES units are located in the distribution network with a total capacity of 1MWh (the size of each CES unit is 50 kWh). Locations of the CES units in the circuit are shown in Figure 8. To provide more potential profit, the CES units are located with the lightest loaded transformers.

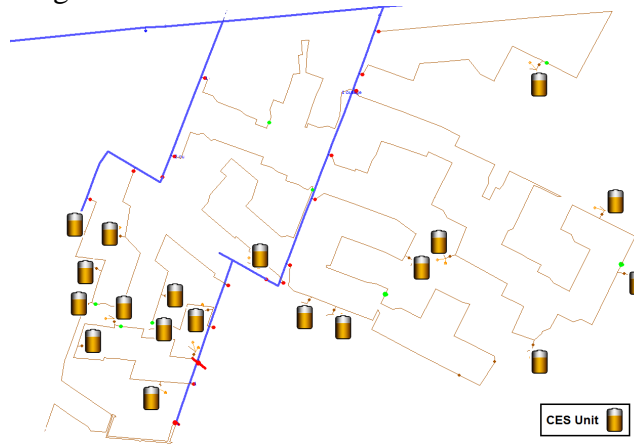

**Figure 8. The case study circuit schematic**

Figure 9 depicts the impact of the aggregated 20 CES units on the feeder loading for a sample day in July. It shows the consistency of CES scheduling with the real time LMP price. During low LMP price hours (6AM - 9 AM), CES units charge to store energy. At the peak hour with expensive LMP prices, CES units discharge to make a profit and decrease the peak load.

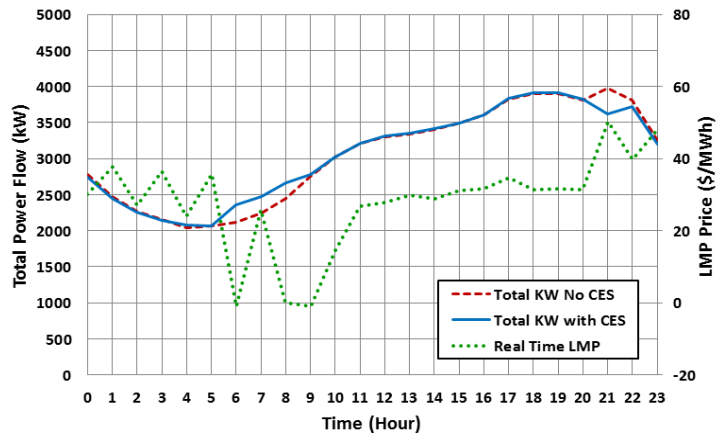

**Figure 9. Feeder load profile for aggregated 20 CES units.**

# 7. Conclusions

A real-time control strategy that maximizes the revenue attainable by community energy storage (CES) systems without sacrificing the occasional benefits related to improvements in reliability, efficiency and reduction in peak feeder loading has been presented. The Gradient-based, Heuristic Optimization algorithm is used for calculating the optimal charge/discharge schedule. Analysis of some of the parameters and tradeoffs involved in operating distributed energy storage devices is presented. It provides insight into such considerations as inaccuracies in LMP price forecasts, transformer loading, reserve capacity, and feeder losses. The analysis shows that economic benefits are highly dependent on the accuracy of the LMP and load forecasts. In the trade-off between accuracy and computational time, incorporating the feeder losses in the objective function has a rather small impact on the total profit, but a rather large impact on the computational time.

Community Energy Storage (CES) systems offer several benefits to electric power system operation, including load support during outages, improved reliability, service availability, renewable energy dispatchability, and peak shaving. While CES installations help meet power system reliability and capacity requirements, the proposed optimal operating strategy further increases the value of the CES.

# 8. Acknowledgments

The authors wish to thank the Department of Energy and DTE Energy for support of this work on the Smart Grid Demonstration Project (DE-OE0000229) entitled "Detroit Edison's Advanced Implementation of Community Energy Storage Systems for Grid Support".